\documentclass[10pt]{iopart}
\usepackage{graphicx} %to allow for width control
\begin{document}

\title[On the vibrational free energy of proteins]{On the vibrational free energy of hydrated proteins}
\author{Yves-Henri Sanejouand}

\address{UFIP, UMR CNRS 6286, Universit{\'{e}} de Nantes, France}
\ead{yves-henri.sanejouand@univ-nantes.fr}
\vspace{10pt}
\begin{indented}
\item[]20 October 2020
\end{indented}

\begin{abstract}
When the hydration shell of a protein is filled with at least 0.6 gram of water per gram of protein, a significant anti-correlation between the vibrational free energy and the potential energy of energy-minimized conformers is observed. 
This means that low potential energy, well-hydrated, protein conformers tend to be more rigid than high-energy ones.
On the other hand, in the case of CASP target 624, when its hydration shell is filled, a significant average energy gap is observed between the crystal structure and the best conformers proposed during the prediction experiment, strongly suggesting that including explicit water molecules may help identifying unlikely conformers among good-looking ones.
\end{abstract}

\vspace{2pc}

\submitto{\PB}
\maketitle
\ioptwocol

\section{Introduction}

The protein folding problem \cite{Wolynes:05,Bowie:05,Dill:12} is important both on a practical and on a theoretical side. On a practical side, solving this problem would allow us to determine structures that can not be obtained either by X-ray crystallography, nuclear magnetic resonance or cryoelectron microscopy. On a theoretical side, it would definitely demonstrate our ability to model the free energy surface of a protein accurately enough, at least as far as the likelihood of its major conformers is concerned.  

Some impressive successes have already been reported. For instance, during the 
13$^{th}$ CASP experiment, in the case of the best model proposed for a 354 residue domain of a xylan acetyltransferase with no known homologue in the PDB, most of the structure core was modelled to a C$_\alpha$ accuracy of better than 2 {\AA} \cite{Casp13}. On the other hand, using standard classical forcefields and straightforward, sub-millisecond long molecular dynamics simulations, the group of David Shaw was able to fold 12 
small fast-folding proteins with, for half of them, a C$_\alpha$ accuracy of better than 1.5 {\AA} \cite{Shaw:11}. 

However, with the latter, so-called \textit{ab initio} approach \cite{SchultenL:02,Dhingra:20}, structural stability problems started to pop up when larger proteins were considered \cite{Shaw:12a}, casting doubts on the accuracy of the underlying potential energy surface.  

Such problems could be due to the specific forcefield used in this later work, namely, the popular CHARMM22* forcefield \cite{Charmm,Shaw:11b}. 
However, they could also come from the too crude description of electrostatical interactions assumed in most classical forcefields used to study macromolecular systems with, in particular, no explicit contribution of atomic and bond polarizabilities \cite{Shaw:19}. Also, standard water models could prove too simple to allow both for a correct behaviour of bulk water and a correct description of water-protein interactions \cite{Pande:05}. 

Work is indeed in progress along these lines. For instance, AMOEBA \cite{Ponder:10}, a polarizable force field, has recently been parallelized \cite{Tinkerhp}. On the other hand, TIP4P-D, a new water model \cite{Shaw:15}, has been developed, the Amber 99SB-ILDN protein forcefield being optimized accordingly, allowing to substantially improve on the state-of-the-art accuracy for simulations of disordered proteins without sacrificing accuracy for folded proteins \cite{Shaw:18}.

In the present study, it is however assumed that such efforts will not prove enough, and this because of another neglected term, namely, the vibrational free energy of the system which, in principle, should be taken into account for each point of the potential energy surface. 

A number of studies have already pointed out that this term may prove important, noteworthy for determining thermodynamical quantities involved in protein folding \cite{Nussinov:00} or in protein-protein \cite{Karplus:94,Mancera:04,Karplus:05p,Gohlke:17} and protein-ligand \cite{Fischer:99,Fischer:01,Smith:04} recognition processes.

The present study confirms that this term can indeed have a significant contribution, at least when the whole hydration shell of the protein is considered. In particular, if it were taken into account, low potential energy conformers would be less likely.  

\section{Methods}

\subsection{Molecular dynamics simulations}

\label{sec:md}
In order to sample the configurational space, molecular dynamics (MD) simulations were performed with Gromacs \cite{Gromacs} version 4.6.3, using the Amber 99SB-ILDN forcefield \cite{Shaw:10amb} and the TIP3P water model \cite{Jorgensen:83}. 
Short range electrostatic and van der Waals interactions were cut off at 12 {\AA}, long-range electrostatics being handled through the particle mesh Ewald method \cite{Darden:93}.

Each considered protein was embedded in a water box, its boundaries being at least 10 {\AA} away from the protein. Sodium and chloride ions were added so as to neutralize the charge of the system and to reach a salt concentration of 150 mM/L. 

The system was then relaxed using steepest descent minimization, with harmonic restraints (force constant of 1,000 kJ/mol/nm ) on protein heavy atoms, until a maximum force threshold of 1,000 kJ/mol/nm was reached. 
The solvent was next equilibrated at 300$^\circ$K, first during 500 ps with a control of both volume and temperature, using Berendsen thermostat  \cite{Berendsen:84} and a coupling constant of 0.1 ps, then during 500 ps with a control of both pressure and temperature, using Parrinello-Rahman barostat  \cite{Parrinello:81} and a coupling constant of 2 ps. 
Finally, restrains were removed and 100 ns of simulation at 300$^\circ$K were
performed, with a timestep of 2 fs, all bonds being constrained with the LINCS
algorithm \cite{LINCS}.

\begin{figure*}[]
\caption{Vibrational free energy, as a function of the potential energy of 100 energy-minimized BPTI conformers. a) When the energy-minimization is performed \textit{in vacuo}. b) When the energy-minimization is performed with the 1,000 closest water molecules. Dashed lines: linear fit.}
\vskip 2mm
\includegraphics[width=\textwidth]{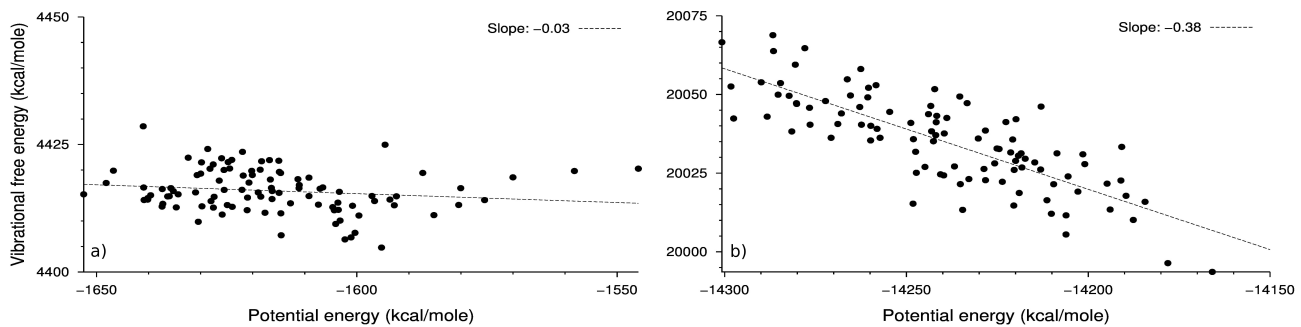}
\label{fig:bpti_corr}
\end{figure*}

\subsection{Vibrational free energy calculations}

\label{sec:vfe}
From each simulation, 100 snapshots were picked (one per nanosecond), the $n_{clos}$ closest water molecules were retained ($n_{clos}$ ranging between 0 and 1,000) and the corresponding hydrated conformers were energy-minimized with Tinker \cite{Tinker8} version 6.2, using the Amber 99SB forcefield, L-BFGS minimization being performed until a gradient of 0.001 kcal/mole/{\AA} was reached (with the minimize program), followed by truncated Newton minimization down to a gradient of 0.00001 kcal/mole/{\AA} (with the newton program), such a strict convergence criterion being required in order to obtain the expected six zero-frequencies \cite{Goldstein:50,Wilson:55,Sanejouand:Phd}.
 
For each energy-minimized conformer, normal mode frequencies, $\nu_i$ being the $i^{th}$ one, were then obtained by diagonalizing the Hessian matrix (with the vibrate program), allowing to obtain $A_{vib}$, the vibrational free energy, which is such that \cite{Karplus:94}: 
\begin{equation}
A_{vib} = \sum^{n_{dof}} \frac{1}{2} h \nu_i + \sum^{n_{dof}} k T \ln( 1 - e^{-\frac{h \nu_i}{k T}} )
\label{eq:avib}
\end{equation}
where $n_{dof}$ is the number of degrees of freedom ($n_{dof} = 3N-6$, $N$ being the number of atoms of the system), $h$ and $k$, respectively, the Planck and Boltzmann constants, 
$T$, the temperature, being set to 300$^\circ$K.

In several previous studies (\textit{e.g.} \cite{Mancera:04,Gohlke:17,Fischer:99,Rios:10}), only
the contribution of the vibrational entropy to $A_{vib}$ was considered. It is given by \cite{Karplus:94}:
\begin{equation}
-T S_{vib} = \sum^{n_{dof}} k T \ln( 1 - e^{-\frac{h \nu_i}{k T}} ) - \sum^{n_{ddl}} \frac{h \nu_i }{e^{\frac{h \nu_i}{k T}} - 1}
\label{eq:svib}
\end{equation}
while the vibrational enthalpy is a follows:
\begin{equation}
H_{vib} = \sum^{n_{dof}} \frac{1}{2} h \nu_i + \sum^{n_{ddl}} \frac{h \nu_i }{e^{\frac{h \nu_i}{k T}} - 1} 
\label{eq:hvib}
\end{equation}
the first sum corresponding to the zero-point energy.

\begin{figure}[b]
\caption{Cumulative energy difference between the highest and the lowest vibrational free energy BPTI conformers, as a function of the mode mean frequency of both conformers, when their 1,000 closest water molecules are taken into account. Plain line: vibrational free energy. Dashed line: contribution of the vibrational entropy ($-T \Delta S_{vib}$). Dotted line: zero-point energy.}
\vskip 2mm
\hskip 3mm
\includegraphics[width=8cm]{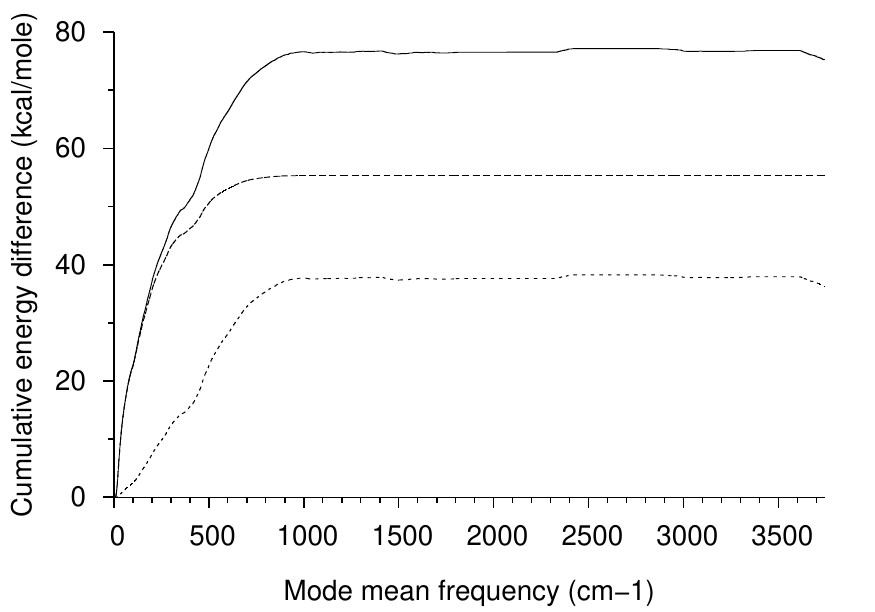}
\label{fig:bpti_diff}
\end{figure}

\section{Results}

\subsection{BPTI conformers}

As shown in Figure \ref{fig:bpti_corr}a, when 100 conformers of the bovine pancreatic trypsin inhibitor (BPTI) picked from a 100 ns MD trajectory (see section \ref{sec:md}) are energy-minimized, the standard deviation of their vibrational free energy (eqn \ref{eq:avib}) is much lower (4.2 kcal/mole) than the standard deviation of their potential energy (18.9 kcal/mole). However, when the energy of each BPTI conformer is minimized together with the 1,000 water molecules that are the closest to this conformer, a significant anti-correlation (-0.79) between the vibrational free energy and the potential energy of the energy-minimized conformers is observed (see Figure \ref{fig:bpti_corr}b), meaning that low-energy hydrated conformers of BPTI are on average significantly more rigid than high energy ones. 

In agreement with previous studies \cite{Karplus:94,Mancera:04,Fischer:99}, and as illustrated in Figure \ref{fig:bpti_diff} by comparing the highest (20,069 kcal/mole) and the lowest vibrational free energy (19,994 kcal/mole) conformers of hydrated BPTI, most of the variation of the vibrational free energy is due to modes with frequencies below 1,000 cm$^{-1}$. Note however that in this case, while the contribution of the vibrational entropy (eqn \ref{eq:svib}) is indeed dominant, the contribution of the vibrational enthalpy (eqn \ref{eq:hvib}) is far from being negligible (26\% of the difference), coming mostly from modes with frequencies ranging between 500 cm$^{-1}$ and 1,000 cm$^{-1}$. 

On the other hand, the zero-point energy (eqn \ref{eq:hvib})
represents 48\% of the vibrational free energy difference between these two conformers (75 kcal/mole), in line with the hypothesis that this term is a key component of the energy of macromolecular systems \cite{Durup:91,Roitberg:95}, like in the case of small molecules \cite{Trinquier:90,Villa:00,Martin:07}.

\begin{figure}[]
\caption{Protein considered. Top left: bovine pancreatic trypsin inhibitor (PDB 4PTI \cite{BPTI}). Top right: ubiquitin (PDB 4XOF \cite{Ubiquitin}). Center: tryptophan-cage (PDB 2JOF \cite{Trpcage}). Bottom left: thioredoxin (PDB 1ERT \cite{Thioredoxin}). Bottom right: hen egg-white lysozyme (PDB 2VB1 \cite{Lysozyme}). Drawn with Chimera \cite{Chimera}.}
\vskip 2mm
\hskip 3mm
\includegraphics[width=8cm]{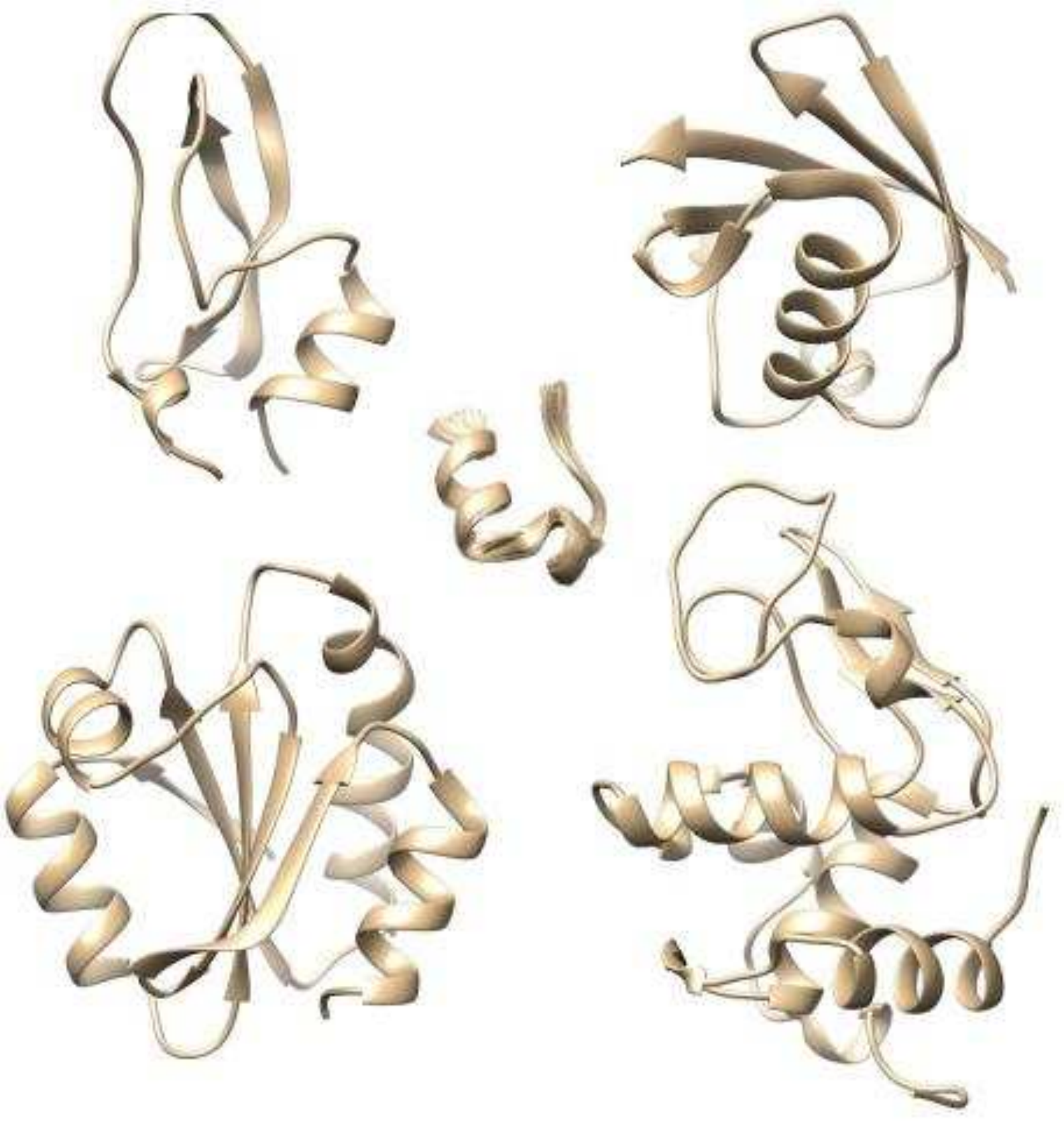}
\label{fig:5cas}
\end{figure}

\subsection{Water hydration threshold}

In order to assess the importance of the anti-correlation between the vibrational free energy and the potential energy of energy-minimized conformers, a least-square fit of the data was performed, by assuming that:
\begin{equation*}
A_{vib} \approx a V_{min} + b
\end{equation*}
where $V_{min}$ is the minimized potential energy of the considered conformer, $a$ being the slope of the vibrational free energy -- potential energy relationship.  

\begin{figure}[]
\caption{Slope of the vibrational free energy -- potential energy linear relationship, as a function of the number of close water molecules taken into account. The vertical (dashed) line corresponds to 0.6 gram of water per gram of protein. Filled circles: BPTI. Filled diamonds: ubiquitin. Stars: tryptophan-cage. Open circles: thioredoxin. Open diamonds: lysozyme.}
\vskip 2mm
\hskip -2mm
\includegraphics[width=9cm]{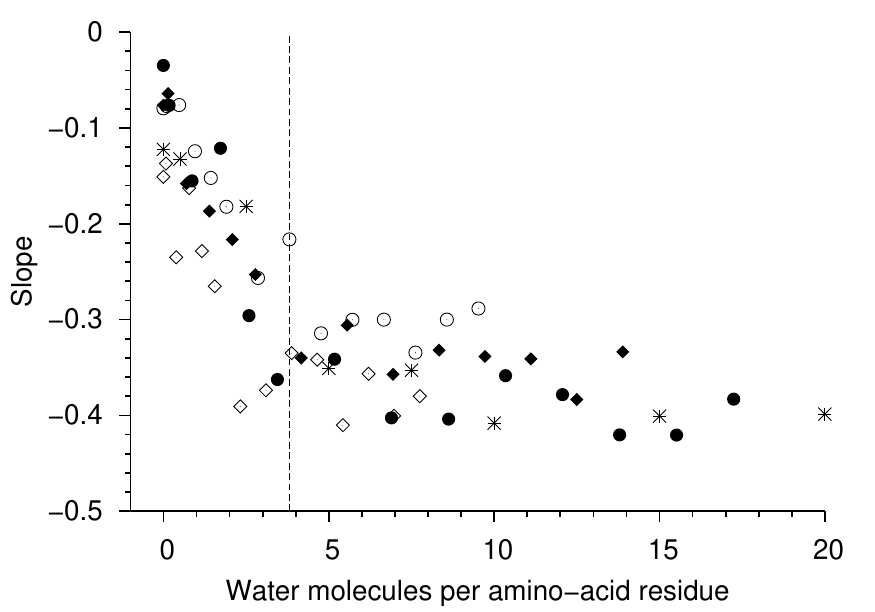}
\label{fig:5cas_slope}
\end{figure}

Together with the BPTI, four other proteins of various folds and sizes were studied (see Figure \ref{fig:5cas}), retaining the 0, 10, 50, 100, 150, 200, 300, 400, 500, 600, 700, 800, 900 or 1,000 closest water molecules of each MD conformer for the analysis (see section \ref{sec:vfe}). 

As shown in Figure \ref{fig:5cas_slope}, like for BPTI (Fig.~\ref{fig:bpti_corr}a), when no water molecule is retained, the slope of the vibrational free energy -- potential energy relationship has a small value, ranging between -0.03, for BPTI, and -0.15, for the lysozyme. On the other hand, when there is more than 3.8 water molecules per amino-acid residue in the considered system, the slope is always below -0.3, confirming that low-energy conformers are on average significantly more rigid than high-energy ones. Moreover, the fact that the value of the slope does not seem to vary significantly when there is more than 3.8 water molecules per residue suggests that this rigidity is not due to the network of hydrogen bonds established between water molecules, but instead to the interactions between the protein and the water molecules belonging to its hydration shell.

Interestingly, 3.8 water molecules per residue corresponds to 0.6 gram of water per gram of protein, that is, the order of magnitude of the amount of tightly bound water molecules found in the hydration shell of proteins \cite{Alary:93} by various techniques like calorimetry \cite{Rupley:83} or microwave dielectric measurements \cite{Buchanan:52,Kodama:96}, in line with the idea that such water molecules are structural ones, that is, that they actually belong to the protein structure \cite{Helms:07}, even though their mean residence time can be quite short \cite{Garcia:93,Halle:03,Laage:12}, namely, on the sub-nanosecond time-scale, except a few notable exceptions \cite{Halle:95,Sanejouand:13}.  

\subsection{Conformers of CASP target 624}

\begin{figure}[]
\caption{Models proposed for CASP target 624. a) ts172-2, b) ts172-3, c) ts172-5, d) ts172-1, e) crystal structure (chain A of PDB 3NRL), f) ts172-4, g) ts236-2, h) ts484-4, i) ts345-1. Drawn with Chimera \cite{Chimera}.}
\vskip 2mm
\hskip 3mm
\includegraphics[width=8cm]{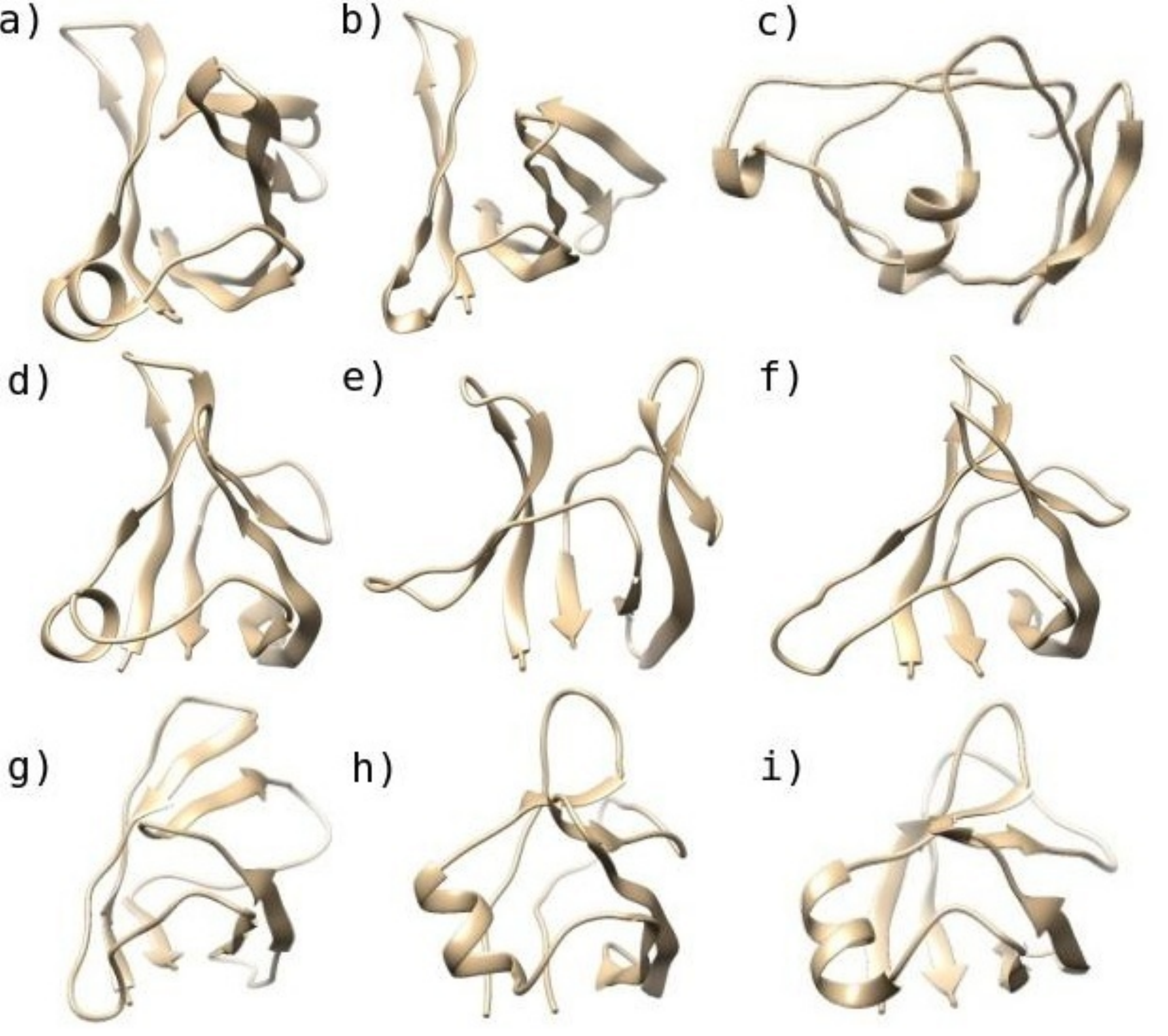}
\label{fig:t624_9cas}
\end{figure}

In order to assess if the vibrational free energy could prove helpful for pinpointing physically relevant protein conformers, eight models proposed for target 624 during the 9$^{th}$ CASP experiment \cite{Casp9} were analysed as above, together with the actual crystallographic structure (see Figure ~\ref{fig:t624_9cas}). Target 624 was chosen because it proved possible to refine the two best models proposed (ts172-1 and ts172-4) through MD simulations on the sub-millisecond time-scale, the C$_\alpha$ root-mean-square (RMSD) to the native structure decreasing from $\approx$ 5 {\AA} down to about 1 {\AA} \cite{Shaw:12a}. 

\begin{figure}[]
\caption{C$_\alpha$ root-mean-square displacement, as a function of time, of nine models of CASP target 624. Filled diamonds: crystal structure. Filled squares: ts172-1. Open squares: ts172-2. Open circles: ts172-3. Filled circles: ts172-4. Open diamonds: ts172-5. Pluses: ts236-2. Crosses: ts345-1. Stars: ts484-4.}
\vskip 2mm
\hskip -2mm
\includegraphics[width=9cm]{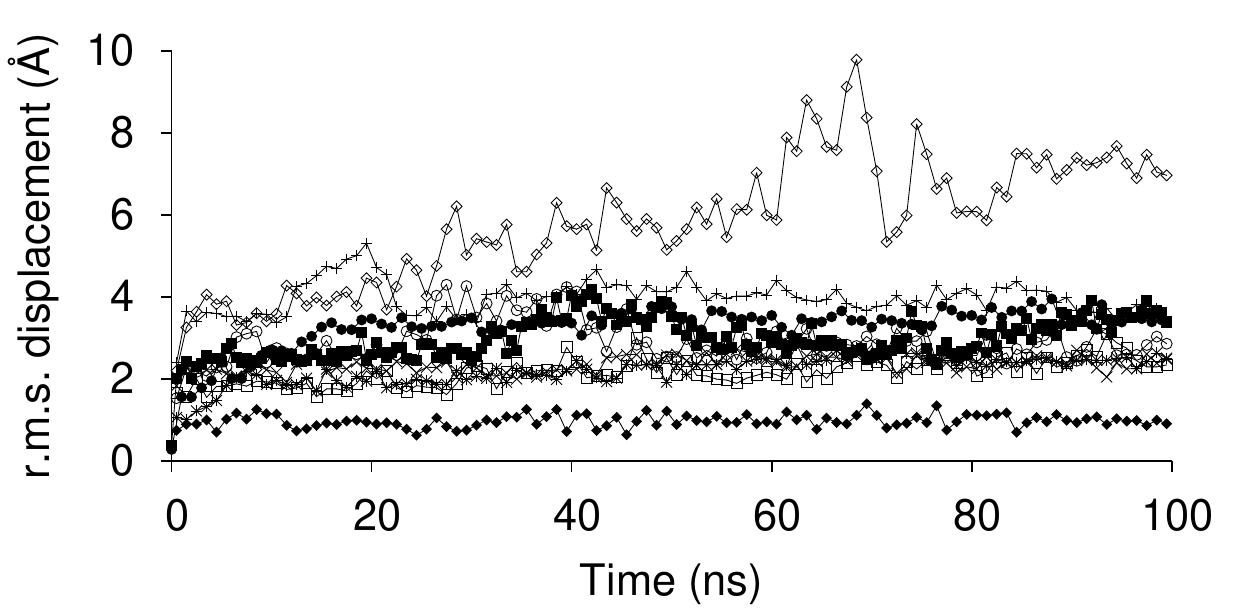}
\label{fig:t624_rmst}
\end{figure}

The five models proposed by the group of David Baker (group 172) were included in the present analysis, as well as three models with a RMSD to the native structure of around 7 {\AA}, proposed by three other groups (ts236-2, ts345-1, ts484-4).
As shown in Figure \ref{fig:t624_rmst}, except ts172-5, all proposed models remain relatively stable during a 100 ns MD simulation, with a RMSD to the initial structure ranging between 2 and 4 {\AA}. Note however that the crystal structure is, according to this criterion, obviously the more stable one, with a RMSD staying below 1.5 {\AA} during the whole simulation. 

\begin{figure}[]
\caption{Average energy of eight models of CASP target 624, as a function of the number of close water molecules taken into account. a) average potential energies obtained after energy-minimization. b) average potential energies + vibrational free energies (VFE). 
The zero (lower dashed line) is given by the average value obtained starting from the crystal structure. The upper dashed line corresponds to 25 kcal/mole more.  
Each average (horizontal bar) is obtained with 100 energy-minimized MD conformers. Averages over 100 kcal/mole are not shown.}
\vskip 2mm
\hskip -2mm
\includegraphics[width=9cm]{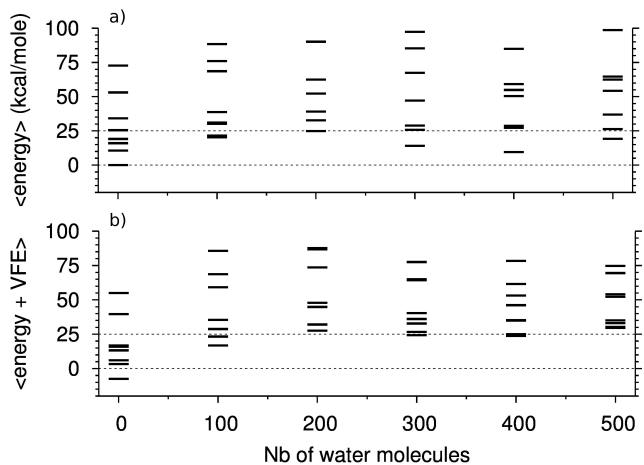}
\label{fig:t624_spect}
\end{figure}

As shown in Figure \ref{fig:t624_spect}, when no water molecule is retained (left column), several models have an average energy similar to, or even better than the average energy obtained starting from the crystal structure, namely: $\langle V_{min} \rangle = -0.01$ kcal/mole, for ts172-1, $\langle V_{min} + A_{vib} \rangle$ = -7.5 kcal/mole, also for ts172-1, +3.3 kcal/mole, for ts345-1, +6.1 kcal/mole, for ts172-2, the averages being obtained with 100 energy-minimized MD conformers (section \ref{sec:vfe}), the zero corresponding to the average value obtained with the crystal structure. 

When water molecules close to the conformers are included in the analysis, the gap between the average energy of the proposed models and the average energy obtained starting from the crystal structure increases dramatically, being around 25 kcal/mole when the vibrational free energy is considered on top of the potential energy, when $n_{clos} \geq$ 200 (see Figure~\ref{fig:t624_spect}b). However, when only the potential energy is taken into account, the gap remains large (see Figure~\ref{fig:t624_spect}a), the smallest values being $\langle V_{min} \rangle$ = +9.5 kcal/mole, for ts484-4 (with $n_{clos}$= 400) and +14.1 kcal/mole, also for ts484-4 (with $n_{clos}$= 300). 

\section{Conclusion}

When the hydration shell of a protein is filled with at least 0.6 gram of water per gram of protein, a significant anti-correlation between the vibrational free energy and the potential energy of energy-minimized conformers is observed (Fig.~\ref{fig:bpti_corr}b), the slope of the corresponding linear relationship ranging between -0.3 and -0.4 (Fig.~\ref{fig:5cas_slope}), depending upon the protein and the number of close water molecules taken into account.

This means that low-energy, well-hydrated, conformers tend to be more rigid than high-energy ones. This also means that if the vibrational free energy of each point of the potential energy surface had been taken into account during the MD simulations, the low potential energy conformers, being less likely, would have been observed less frequently. 

On the other hand, for CASP target 624, 
when its hydration shell is filled, a significant energy gap is observed between the crystal structure and the best conformers that were proposed during the prediction experiment (Fig.~\ref{fig:t624_spect}), strongly suggesting that including explicit water molecules may help identifying unlikely conformers  among good-looking ones.
However, when the vibrational free energy is added on top of the potential energy, 
the gap increases in a few cases \textit{only}, meaning that, for pinpointing  unlikely conformers, the average potential energy of energy-minimized, well hydrated, conformers is a powerful enough criterion, at least in the case of target 624. 

The present work rely on the hypothesis that studying a sample of potential energy minima, that is, the inherent structures of the system \cite{Stillinger:97}, can yield a fair approximation for its average vibrational energy. Though the results thus obtained look promising, such an hypothesis can not be taken for granted. A possible way to check it would be to take advantage of the fact that vibrational free energies can be obtained from the vibrational density of states function \cite{Ochsenfeld:19}, while the latter can for instance be obtained through the Fourier transform of the velocity autocorrelation function. Such an approach could indeed allow to obtain the average vibrational free energy of an hydrated protein \textit{directly} from a standard, room temperature, molecular dynamics simulation \cite{Goddard:03}. 

\section*{Acknowledgements}

I thank Jean Durup for stimulating discussions on this topic, some time ago.

\end{document}